\UseRawInputEncoding
\documentclass[aps, prc, floatfix, nofootinbib, superscriptaddress, twocolumn]{revtex4-1}
\usepackage{latexsym}
\usepackage{amsmath}
\usepackage{amssymb}
\usepackage{amsfonts}
\usepackage{makecell}

\usepackage[mathscr,scaled=1.15]{urwchancal}
\DeclareFontFamily{OT1}{pzc}{}
\DeclareFontShape{OT1}{pzc}{m}{it}%
{<-> s * [1.15] pzcmi7t}{}
\DeclareMathAlphabet{\mathpzc}{OT1}{pzc}{m}{it}

\usepackage{color}

\usepackage{supertabular}
\usepackage{placeins}
\usepackage{epsfig}
\usepackage{graphicx}
\usepackage{multirow}

\definecolor{purple}{rgb}{0.5,0,0.5}
\definecolor{blue}{rgb}{0.0,0,0.9}
\definecolor{prdblue}{rgb}{0.133,0.118,0.498}
\usepackage[colorlinks=true, pdfstartview=FitV, linkcolor=prdblue, citecolor= prdblue, urlcolor=prdblue]{hyperref}

\begin{document}

\title{Double-heavy tetraquarks with strangeness in the chiral quark model}

\author{Xiaoyun Chen}
\email[]{xychen@jit.edu.cn}
\affiliation{College of Science, Jinling Institute of Technology, Nanjing 211169, P. R. China}

\author{Fu-Lai Wang}
\email[]{wangfl2016@lzu.edu.cn}
\affiliation{School of Physical Science and Technology, Lanzhou University, Lanzhou 730000, China}
\affiliation{Research Center for Hadron and CSR Physics, Lanzhou University and Institute of Modern Physics of CAS, Lanzhou 730000, China}

\author{Yue Tan}
\email[]{181001003@njnu.edu.cn}
\affiliation{Department of Physics, Yancheng Institute of Technology, Yancheng 224000, P. R. China}

\author{Youchang Yang}
\email[]{yangyc@gues.edu.cn}
\affiliation{Guizhou University of Engineering Science; Zunyi Normal University, P. R. China}

\begin{abstract}
Recently, some progresses have been made on the double-heavy tetraquarks in the experiments, such as $T_{cc}$ reported by LHCb Collaboration, and $X_{cc\bar{s}\bar{s}}$ reported by the Belle Collaboration. Coming on the heels of our previous work about $T_{cc}$ and $T_{bb}$, we present a study on the bound states and the resonance states of its companions $QQ\bar{q}\bar{s}$ ($Q=c,b; q=u, s$) tetraquarks with strange flavor in the chiral quark model. Two pictures, one with meson-meson picture, another with diquark-antidiquark picture and their couplings are considered in our calculations. Isospin violation is neglected herein. Our numerical analysis indicates that the states $cc\bar{u}\bar{s}$ with $\frac{1}{2}(1^+)$ and $bb\bar{u}\bar{s}$ with $\frac{1}{2}(1^+)$ are the most promising stable states against strong interactions. Besides, we also find several resonance states for the double-heavy strange tetraquarks with the real scaling method.
\end{abstract}

\maketitle

%%%%%%%%%%%%%%%%%%%%%%%%%%%%%%%%%%%%%%%%%%%%%%%%%%%%%%%%%%%%%%%%%%%%%%%%%%%%%%%%%%%%%%%%%%%%%%%%%%%%%%%%%%%%%%%%%%%%%%%

\section{Introduction}
\label{introduction}
Searching for exotic hadrons, made of more than three quarks(antiquarks), is one of the most interesting subjects in hadron physics, because they may contain abundant low-energy strong interaction information than ordinary hadrons. A large amount of new hadrons have been observed by experiments since the Belle collaboration’s discovery of the state $X(3872)$ in 2003 \cite{Belle:2003nnu}. For these new observed hadron states, some of them cannot be explained as the conventional mesons because they are charged states, such as $Z_c^+$, and some others cannot be explained very well in the traditional $qq\mbox{-}$meson or $qqq\mbox{-}$baryon framework. Multiquark states have therefore attracted much attention theoretically.

Recent experimental studies of the heavy meson spectroscopy revealed several new states, such as $T_{cc}$ reported by LHCb Collaboration \cite{LHCb:2021vvq, LHCb:2021auc}, and $X_{cc\bar{s}\bar{s}}$ by the Belle Collaboration \cite{Belle:2021kub}, which cannot be simply accommodated in the quark-antiquark ($q\bar{q}$) picture \cite{Swanson:2006st}. These states may be good candidates of multiquark states. The simplest multiquark system is a tetraquark, made up of two quarks and two antiquarks. Heavy tetraquarks are of particular interest, because the presence of a heavy quark will increase the binding energy of the bound system. As a consequence, such tetraquarks will have masses below the thresholds and forbid the strong decays to mesons with open heavy flavor, with the small decay width generated by weak and electromagnetic decay. So it is important to investigate the possible stability of the $QQ\bar{q}\bar{q}$ tetraquarks since they are explicitly exotic states with the heavy flavor number equal to 2. And their observations will be a direct proof of the existence of multiquark states.

Until now, there are many theoretical studies about the double-heavy tetraquarks in various phenomenological methods, such as constituent quark model, chiral perturbation theory, string model, lattice QCD and QCD sum rule approach \cite{Eichmann:2020oqt,Zhang:2007mu,Deng:2018kly,Eichten:2017ffp,Francis:2016hui,Cheng:2020wxa,Qin:2020zlg,Park:2018wjk,
Silvestre-Brac:1993zem,Meng:2020knc,Meng:2021yjr,Weng:2021hje,Guo:2021yws,Wu:2021rrc,Junnarkar:2018twb,Pflaumer:2021ong,Agaev:2020zag,
Karliner:2021wju,Esposito:2013fma,Ding:2021igr,Braaten:2020nwp,Luo:2017eub,Tang:2019nwv,Zhu:2019iwm,Noh:2021lqs,Ebert:2007rn,Lu:2020rog,Hudspith:2020tdf,Yang:2009zzp,
Yang:2019itm,Du:2012wp}. Most of them predicted that $cc\bar{u}\bar{d}$, $bb\bar{u}\bar{d}$ with $0(1^+)$ and $cc\bar{u}\bar{s}$, $bb\bar{u}\bar{s}$ with $\frac{1}{2}(1^+)$ have the stable masses below the corresponding thresholds. Yet, none of them has so far been confirmed experimentally except $T_{cc}$. what's more, the states with masses above some two-hadron thresholds become resonances with often a large fall-apart decay width, which are also the possibly observable objects experimentally in the future.

Following our previous work about $T_{cc}$ and $T_{bb}$ \cite{Chen:2021tnn}, the purpose of this paper is to systematically study tetraquarks with two heavy quarks $QQ (Q=b,c)$ and two light/strange quarks $us$ or $ss$ with possible quantum numbers constrained by the Pauli principle, and try to find stable bound states and resonance states in the strong interaction in the non-relativistic chiral quark model.
According to previous results about $T_{cc}$, the different quark picture can lead to diverse conclusions. For instance, the diquark-antidiquark structure leads to the deeply bound states, whereas the meson-meson picture brings about weakly bound state. When considering the coupling of the two pictures, we can get the deeper bound states, but the weakly bound state disappears. We may look forward to the future more precise experimental data if there exist states with much deeper binding energy.

What's more, along with the discoveries of the new hadron states in experiments, we are blessed with new opportunities and faced with new challenges, such as, the chiral quark model. In the quark model, the quark-quark interactions have a good achievement on hadron spectrum, where the unique color structure, color-singlet, is accepted. Nowadays, we usually generalize the quark-quark interactions used in color-singlet baryons and mesons to multiquark systems, such as using Casimir scaling ~\cite{Casimir}. With the accumulations of the experimental data, it can deepen our understanding of the hadron-hadron interactions and may shed light on the properties of the newly observed exotic states.

In our present work, different quark configurations such as the meson-meson picture, the diquark-antidiquark picture and their couplings are also considered to show the stability of states. For $T_{cc}$ states, the $SU(2)$ flavor symmetry is applied, and for the double-heavy tetraquarks with strangeness, $SU(2)$ flavor symmetry may be a good choice, but the Goldstone boson $K$ meson exchanges will be inoperative. So for the strange double-heavy tetraquarks, in the introducing of the Hamiltonian for the hadron states, the octet Goldstone boson exchanges $V^{\pi, K, \eta}$ are considered. But the $SU(3)$ flavor symmetry breaking is also taken into account, with the different masses of $u,d$ and $s$ quark, and gives rise to the different masses of $\pi$ and $K$. In the construction of the wavefunctions, we also takes the mass differences into consideration.

We find no bound stats for strange double-heavy tetraquarks when pure meson-meson or diquark-antidiquark pictures are considered, but we find a bound state for $cc\bar{u}\bar{s}$ and $bb\bar{u}\bar{s}$ with isospin and spin-parity $I(J^P)=\frac{1}{2}(1^+)$ only when considering the coupling of the meson-meson and diquark-antidiquark pictures, having a binding energy of 7.0 MeV and 104.4 MeV, respectively. And for $bb\bar{u}\bar{s}$ with $I(J^P)=\frac{1}{2}(1^+)$, the $K$ meson exchange is much larger than the $cc\bar{u}\bar{s}$ system and plays a role in obtaining the deeper bound energy. Meanwhile, the possible resonance states are also searched within a real scaling method (RSM) \cite{rs1} in the full channel-coupling calculations. With the help of RSM, we obtained some observable resonances for double-heavy tetraquarks with strangeness.

The article is organized as follows. In the next section we perform the framework of the chiral quark model and the wave functions of the tetraquark states. In Sec. \ref{results}, we present analyses of our results. We conclude with a summary and outlook in Sec. \ref{summary}.

%%%%%%%%%%%%%%%%%%%%%%%%%%%%%%%%%%%%%%%%%%%%%%%%%%%%%%%%%%%%%%%%%%%%%%%%%%%%%%%%%%%%%%%%%%%%%%%%%%%%%%%%%%%%%%%%%%%%%%%

\section{Formalism}
\label{Formalism}
\subsection{Hamiltonian}
As we all know, quantum Chromodynamics (QCD) has three important properties, asymptotic freedom, color confinement, approximate chiral symmetry and its spontaneous breaking. Until now, it is still difficult for us to obtain the hadron spectra analytically from the QCD Lagrangian. The QCD-inspired chiral quark model is widely regarded as an effective tool to get the hadron spectra and understand the hadron-hadron interactions. In the chiral quark model, the quark-quark interactions within confinement scale ($\sim$ 1 fm) have undergone a wide check in the hadron spectra, where the unique color structure, color-singlet, is accepted.
For the multiquark system, with various color structures, the effects of color structures are considered by Casimir operator
$\boldsymbol{\lambda}_i^c\cdot \boldsymbol{\lambda}_j^c$ \cite{Bali:2000un}.

In the chiral quark model, the Hamiltonian of the four-quark system reads
\begin{align}\label{Hm}
 H & = \sum_{i=1}^4 m_i  +\frac{\bf{p^2_{12}}}{2\mu_{12}}+\frac{\bf{p^2_{34}}}{2\mu_{34}}
  +\frac{\bf{p^2_{r}}}{2\mu_{r}}  \quad  \nonumber \\
  & + \sum_{i<j=1}^4 \left[ V_{ij}^{C}+V_{ij}^{G}+\sum_{\chi=\pi,K,\eta} V_{ij}^{\chi}
   +V_{ij}^{\sigma}\right],
\end{align}
where $m_i$ is the constituent mass of $i$th quark (antiquark). $\frac{\bf{p^2_{ij}}}{2\mu_{ij}} (ij=12; 34)$ and $\frac{\bf{p^2_{r}}}{2\mu_{r}}$ represents the inner kinetic of two-cluster and the relative motion kinetic between two clusters, respectively, with
\begin{subequations}
\begin{align}
\bf{p}_{12}&=\frac{m_2\mathbf{p}_1-m_1\mathbf{p}_2}{m_1+m_2}, \\
\mathbf{p}_{34}&=\frac{m_4\mathbf{p}_3-m_3\mathbf{p}_4}{m_3+m_4},  \\
\mathbf{p}_{r}&= \frac{(m_3+m_4)\mathbf{p}_{12}-(m_1+m_2)\mathbf{p}_{34}}{m_1+m_2+m_3+m_4}, \\
\mu_{ij}&=\frac{m_im_j}{m_i+m_j}, \\
\mu_{r}&=\frac{(m_1+m_2)(m_3+m_4)}{m_1+m_2+m_3+m_4}.
\end{align}
\end{subequations}
The potential energy is constituted from pieces describing quark confinement ``$V_{ij}^C$'', one-gluon-exchange (OGE) ``$V_{ij}^G$'', Goldstone boson exchange ``$V_{ij}^{\chi=\pi, K, \eta}$'' and scalar $\sigma$ meson exchange; and the central part form of them for the four-quark system is \cite{Valcarce:2005em}:
{\allowdisplaybreaks
\begin{subequations}
\begin{align}
V_{ij}^{C}&= ( -a_c r_{ij}^2-\Delta ) \boldsymbol{\lambda}_i^c
\cdot \boldsymbol{\lambda}_j^c ,  \\
 V_{ij}^{G}&= \frac{\alpha_s}{4} \boldsymbol{\lambda}_i^c \cdot \boldsymbol{\lambda}_{j}^c
\left[\frac{1}{r_{ij}}-\frac{2\pi}{3m_im_j}\boldsymbol{\sigma}_i\cdot
\boldsymbol{\sigma}_j
  \delta(\boldsymbol{r}_{ij})\right],  \nonumber\\
\delta{(\boldsymbol{r}_{ij})} & =  \frac{e^{-r_{ij}/r_0(\mu_{ij})}}{4\pi r_{ij}r_0^2(\mu_{ij})}, r_0(\mu_{ij}) =s_0/\mu_{ij}, \\
V_{ij}^{\pi}&= \frac{g_{ch}^2}{4\pi}\frac{m_{\pi}^2}{12m_im_j}
  \frac{\Lambda_{\pi}^2}{\Lambda_{\pi}^2-m_{\pi}^2}m_\pi v_{ij}^{\pi}
  \sum_{a=1}^3 \lambda_i^a \lambda_j^a,  \\
V_{ij}^{K}&= \frac{g_{ch}^2}{4\pi}\frac{m_{K}^2}{12m_im_j}
  \frac{\Lambda_K^2}{\Lambda_K^2-m_{K}^2}m_K v_{ij}^{K}
  \sum_{a=4}^7 \lambda_i^a \lambda_j^a,   \\
\nonumber
V_{ij}^{\eta} & =
\frac{g_{ch}^2}{4\pi}\frac{m_{\eta}^2}{12m_im_j}
\frac{\Lambda_{\eta}^2}{\Lambda_{\eta}^2-m_{\eta}^2}m_{\eta}
v_{ij}^{\eta}  \\
 & \quad \times \left[\lambda_i^8 \lambda_j^8 \cos\theta_p
 - \lambda_i^0 \lambda_j^0 \sin \theta_p \right],   \nonumber \\
 v_{ij}^{\chi} & =  \left[ Y(m_\chi r_{ij})-
\frac{\Lambda_{\chi}^3}{m_{\chi}^3}Y(\Lambda_{\chi} r_{ij})
\right]
\boldsymbol{\sigma}_i \cdot\boldsymbol{\sigma}_j, \\
V_{ij}^{\sigma}&= -\frac{g_{ch}^2}{4\pi}
\frac{\Lambda_{\sigma}^2}{\Lambda_{\sigma}^2-m_{\sigma}^2}m_\sigma \nonumber\\
& \quad \times \left[
 Y(m_\sigma r_{ij})-\frac{\Lambda_{\sigma}}{m_\sigma}Y(\Lambda_{\sigma} r_{ij})\right]  , \nonumber \\
  Y(x)  &=   e^{-x}/x.
\end{align}
\end{subequations}}
\hspace*{-0.5\parindent}%
$\boldsymbol{\sigma}$ are the $SU(2)$ Pauli matrices.
$\boldsymbol{\lambda}$, $\boldsymbol{\lambda}^c$ are the $SU(3)$ flavor, color Gell-Mann matrices, respectively.
$g^2_{ch}/4\pi$ is the chiral coupling constant, determined from the $\pi$-nucleon coupling.
and $\alpha_s$ is an effective scale-dependent running coupling \cite{Valcarce:2005em},
\begin{equation}
\alpha_s(\mu_{ij})=\frac{\alpha_0}{\ln\left[(\mu_{ij}^2+\mu_0^2)/\Lambda_0^2\right]}.
\end{equation}
$\Lambda_0,\alpha_0, \mu_0, s_0$ are adjustable model parameters, and $\theta_p$ is fixed by $\eta$ and $\eta^{'}$ mixing.

It is important to be noted that the noncentral part of the OGE, tensor and spin-orbit forces, between quarks are omitted in the present calculations, since only $S\mbox{-}$wave tetraquarks are studied here, and the contributions of the noncentral part are small or approximatively zero. Secondly, for $QQ\bar{q}\bar{s}~(Q=c, b; q=u, s)$ systems, Goldstone bosons potential $V^{\pi}$ and $V^{\sigma}$ will be zero herein.
Lastly, we show the model parameters \cite{Vijandemodel} in Table \ref{modelparameters}. Need to note that, in the reference \cite{Vijandemodel}, the confinement item takes the form $V^{C}_{ij}=\big(-a_c(1-e^{-\mu_c r_{ij}}\big)+\Delta)(\boldsymbol{\lambda}_i^c \cdot \boldsymbol{\lambda}_j^c)$.
And in our present calculations, the usual quadratic confinement $V^{C}_{ij}=( -a_c r_{ij}^2-\Delta ) \boldsymbol{\lambda}_i^c
\cdot \boldsymbol{\lambda}_j^c$ is employed, so some parameters are different such as quark mass, $a_c$ and $\Delta$.

Using the model parameters, we calculated the ground-state masses of the most mesons, especially the relevant mesons $D$, $D^*$, $D_s$, $D_s^*$, $B$, $B^*$, $B_s$, $B_s^*$ in the present work, which are demonstrated in Table \ref{mesonspectra}. From the table, we can find that the quark model achieves great success on describing the hadron spectra, especially for the ground-state heavy mesons.

\begin{table}[!t]
\begin{center}
\caption{ \label{modelparameters}
Model parameters, determined by fitting the meson spectra.}
\begin{tabular}{llr}
\hline\noalign{\smallskip}
Quark masses   &$m_u=m_d$    &313  \\
   (MeV)       &$m_s$         &536  \\
               &$m_c$         &1728 \\
               &$m_b$         &5112 \\
\hline
Goldstone bosons   &$m_{\pi}$     &0.70  \\
   (fm$^{-1} \sim 200\,$MeV )     &$m_{\sigma}$     &3.42  \\
                   &$m_{\eta}$     &2.77  \\
                   &$m_{K}$     &2.51  \\
                   &$\Lambda_{\pi}=\Lambda_{\sigma}$     &4.2  \\
                   &$\Lambda_{\eta}=\Lambda_{K}$     &5.2  \\
                   \cline{2-3}
                   &$g_{ch}^2/(4\pi)$                &0.54  \\
                   &$\theta_p(^\circ)$                &-15 \\
\hline
Confinement        &$a_c$ (MeV fm$^{-2}$)         &101 \\
                   &$\Delta$ (MeV)     &-78.3 \\
\hline
OGE                 & $\alpha_0$        &3.67 \\
                   &$\Lambda_0({\rm fm}^{-1})$ &0.033 \\
                  &$\mu_0$(MeV)    &36.98 \\
                   &$s_0$(MeV)    &28.17 \\
\hline
\end{tabular}
\end{center}
\end{table}

\linespread{1.2}
\begin{table}[!t]
\begin{center}
\renewcommand\tabcolsep{10.0pt} % 调整表格列间的宽度
\caption{ \label{mesonspectra} The ground-state masses of the mesons in the chiral quark model in comparison with the experimental data \cite{PDG} (in unit of MeV).}
\begin{tabular}{cccc}
\hline\hline\noalign{\smallskip}
Meson &$I(J^P)$  &Mass &PDG \cite{PDG}\\
\hline
$\pi$ &$1(0^-)$ &139.3 &139.6 \\
$K$   &$\frac{1}{2}(0^-)$ &493.9 &493.7 \\
$\rho$ &$1(1^-)$ &772.0 &770.0 \\
$K^*$  &$\frac{1}{2}(1^-)$ &914.0 &892.0\\
$\omega$ &$0(1^-)$ &701.6 &782.7 \\
$\eta$   &$0(0^-)$ &669.2 &547.0\\
$\phi(1020)$   &$0(1^-)$ &1016.5 &1019.4 \\
$\eta_c(1s)$   &$0(0^-)$ &2986.3 &2983.6\\
$J/\Psi$       &$0(1^-)$  &3096.4 &3096.9\\
$D^0$ &$\frac{1}{2}(0^-)$  &1862.6     &1864.8 \\
$D^{*0}$ &$\frac{1}{2}(1^-)$  &1980.6        &2007.0\\
$D_s^+$ &$0(0^-)$  &1950.2     &1968.3 \\
$D_s^{*+}$ &$0(1^-)$  &2079.9        &2112.2\\
$B^-$ &$\frac{1}{2}(0^-)$  &5280.8     &5279.3 \\
$B^{*-}$ &$\frac{1}{2}(1^-)$  &5319.6        &5325.2\\
$\bar{B_s}^0$ &$0(0^-)$  &5367.7     &5366.8 \\
$\bar{B_s^*}^0$ &$0(1^-)$  &5410.2        &5415.4\\
$\eta_b(1s)$ &$0(0^-)$ &9334.7 &9399.1 \\
$\Upsilon(1s)$ &$0(1^-)$ &9463.9  &9460.3\\
\hline\hline
\end{tabular}
\end{center}
\end{table}

\subsection{Wave function}
The properties of the double-heavy tetraquark states can be obtained with a complete wave function, which should be a direct product of the orbital, spin, color and flavor wave functions that contribute to a given well defined quantum numbers $I(J^P)$.

For orbital part, in our calculations, the orbital wave function is
\begin{equation}\label{spatialwavefunctions}
\Psi_{L}^{M_{L}}=\left[[\Psi_{l_1}({\bf r})\Psi_{l_2}({\bf
R})]_{l_{12}}\Psi_{L_r}(\bf{Z}) \right]_{L}^{M_{L}},
\end{equation}
where, $\bf{r}$, $\bf{R}$ and $\bf{Z}$ are the relative spatial coordinates, and one of the definitions of the Jacobi coordinates can be written as,
\begin{align}
\bf{r}&=r_1-r_2, \nonumber \\
\bf{R}&=r_3-r_4, \nonumber \\
\bf{Z}&=\frac{m_1{\bf r}_1+m_2{\bf r}_2}{m_1+m_2}-\frac{m_3{\bf r}_3+m_4{\bf r}_4}{m_3+m_4}, \nonumber \\
\bf{R}_c&=\frac{m_1{\bf r}_1+m_2{\bf r}_2+m_3{\bf r}_3+m_4{\bf r}_4}{m_1+m_2+m_3+m_4}.
\end{align}
$\bf{R}_c$ is the center-of-mass coordinate. In Eq. (\ref{spatialwavefunctions}), $l_1$, $l_2$ is the inner angular momentum of the two sub-cluster; $L_r$ is the relative angular momentum between two sub clusters. $L$ is the total orbital angular momentum of the four-quark system, with $L=l_1 \oplus l_2 \oplus L_r$. In the present work, we just consider the low-lying $S\mbox{-}$wave double heavy tetraquark states, so it is natural to assume that all the orbital angular momenta are zeros. The parity of the double-heavy tetraquarks $QQ\bar{q}\bar{s}$ can be expressed in terms of the relative orbital angular momenta, with $P=(-1)^{l_1+l_2+L_r}=+1$. So as to get the reliable information of the four-quark system, a high precision numerical method, Gaussian expansion method~(GEM) \cite{GEM} is applied in our work. In GEM, any relative motion wave function can be expanded in series of Gaussian basis functions,
\label{radialpart}
\begin{align}
\Psi_{l}^{m}(\mathbf{x}) & = \sum_{n=1}^{n_{\rm max}} c_{n}N_{nl}x^{l}
e^{-\nu_{n}x^2}Y_{lm}(\hat{\mathbf{x}}),
\end{align}
where $N_{nl}$ are normalization constants,
\begin{align}
N_{nl}=\left[\frac{2^{l+2}(2\nu_{n})^{l+\frac{3}{2}}}{\sqrt{\pi}(2l+1)}
\right]^\frac{1}{2}.
\end{align}
$c_n$ are the variational parameters, which are determined dynamically. The Gaussian size
parameters are chosen according to the following geometric progression
\begin{equation}\label{gaussiansize}
\nu_{n}=\frac{1}{r^2_n}, \quad r_n=r_1a^{n-1}, \quad
a=\left(\frac{r_{n_{\rm max}}}{r_1}\right)^{\frac{1}{n_{\rm
max}-1}}.
\end{equation}
This procedure enables optimization of the expansion using just small numbers of Gaussians. For example, in order to obtain the stable ground-state masses of mesons in Table \ref{mesonspectra}, we takes,
\begin{equation}
r_1=0.01~\rm{fm}, \quad \emph{r}_{\emph{n}_{\rm max}}=2~\rm{fm}, \quad \emph{n}_{\rm{max}}=12.
\end{equation}

For the spin part, we first write down the wave functions for two-quark system,
\begin{subequations}
\begin{align}
\chi_{11}&=\alpha\alpha,\\
\chi_{10}&=\frac{1}{\sqrt{2}}(\alpha\beta+\beta\alpha),\\
\chi_{1-1}&=\beta\beta, \\
\chi_{00}&=\frac{1}{\sqrt{2}}(\alpha\beta-\beta\alpha).
\end{align}
\end{subequations}
If the spin of one cluster is coupled  to $S_1$ and that of another cluster to $S_2$, the total spin wave function of the four-quark system can be obtained as $S=S_1 \oplus S_2$,
\begin{subequations}
\begin{align}
\chi_{00=0\oplus0}^{\sigma1}&=\chi_{00}\chi_{00}, \\
\chi_{00=1\oplus1}^{\sigma2}&=\sqrt{\frac{1}{3}}(\chi_{11}\chi_{1-1}-\chi_{10}\chi_{10}+\chi_{1-1}\chi_{11}), \\
\chi_{11=0\oplus1}^{\sigma3}&=\chi_{00}\chi_{11}, \\
\chi_{11=1\oplus0}^{\sigma4}&=\chi_{11}\chi_{00},\\
\chi_{11=1\oplus1}^{\sigma5}&=\frac{1}{\sqrt{2}}(\chi_{11}\chi_{10}-\chi_{10}\chi_{11}),\\
\chi_{22=1\oplus1}^{\sigma6}&=\chi_{11}\chi_{11}.
\end{align}
\end{subequations}
The subscript of $\chi$ represents the  $SM_S=S_1 \oplus S_2$, and $M_S$ is the third projection of the total spin $S$.

For the meson-meson picture, the indices of particles are ``1234'', and for the diquark-antidiquark picture, the indices are ``1324''.
For the color part, in the meson-meson picture, the colorless wave functions can be obtained from $\big[[Q\bar{q}]_{1_c}[Q\bar{s}]_{1_c}\big]_1$ or $\big[[Q\bar{q}]_{8_c}[Q\bar{s}]_{8_c}\big]_1$. In the diquark-antidiquark picture, the color representation of the diquark maybe antisymmetrical $[QQ]_{\bar{3}_c}$ or symmetrical $[QQ]_{6_c}$, and for antidiquark, the color form is antisymmetrical $[\bar{q}\bar{s}]_{3_c}$ or symmetrical $[\bar{q}\bar{s}]_{\bar{6}_c}$. There are two rules to couple the diquark and antidiquark into a colorless wave function: one is the good diquark with attractive interaction $\big[[QQ]_{\bar{3}_c}[\bar{q}\bar{s}]_{3_c}\big]_1$, and another is the bad diquark with repulsive interaction $\big[[QQ]_{6_c}[\bar{q}\bar{s}]_{\bar{6}_c}\big]_1$.
So we can easily write down the color wave functions in the meson-meson picture and the diquark-antidiquark picture, respectively.
\begin{subequations}
\begin{align}
\chi^{c1}_{1 \otimes 1}&=\frac{1}{3}(r\bar{r}+g\bar{g}+b\bar{b})_{12}(r\bar{r}+g\bar{g}+b\bar{b})_{34}, \\
\chi^{c2}_{8 \otimes 8}&=\frac{\sqrt{2}}{12}(3r\bar{b}b\bar{r}+3r\bar{g}g\bar{r}+3g\bar{b}b\bar{g}+3b\bar{g}g\bar{b} \nonumber \\
&+3g\bar{r}r\bar{g}+3b\bar{r}r\bar{b}+2r\bar{r}r\bar{r}+2g\bar{g}g\bar{g} \nonumber \\
&+2b\bar{b}b\bar{b}-r\bar{r}g\bar{g}-g\bar{g}r\bar{r}-b\bar{b}g\bar{g} \nonumber \\
&-b\bar{b}r\bar{r}-g\bar{g}b\bar{b}-r\bar{r}b\bar{b})_{1234}, \\
\chi^{c3}_{\bar{3} \otimes 3}&=\frac{\sqrt{3}}{6}(rg\bar{r}\bar{g}-rg\bar{g}\bar{r}+gr\bar{g}\bar{r}-gr\bar{r}\bar{g} \nonumber \\
&+rb\bar{r}\bar{b}-rb\bar{b}\bar{r}+br\bar{b}\bar{r}-br\bar{r}\bar{b} \nonumber \\
&+gb\bar{g}\bar{b}-gb\bar{b}\bar{g}+bg\bar{b}\bar{g}-bg\bar{g}\bar{b})_{1324},  \\
\chi^{c4}_{6 \otimes \bar{6}}&=\frac{\sqrt{6}}{12}(2rr\bar{r}\bar{r}+2gg\bar{g}\bar{g}+2bb\bar{b}\bar{b}+rg\bar{r}\bar{g}\nonumber \\
&+rg\bar{g}\bar{r}+gr\bar{g}\bar{r}+gr\bar{r}\bar{g}+rb\bar{r}\bar{b}\nonumber \\
&+rb\bar{b}\bar{r}+br\bar{b}\bar{r}+br\bar{r}\bar{b}+gb\bar{g}\bar{b}\nonumber \\
&+gb\bar{b}\bar{g}+bg\bar{b}\bar{g}+bg\bar{g}\bar{b})_{1324}.
\end{align}
\end{subequations}

For the flavor wave functions of $QQ\bar{q}\bar{s}~(Q=c, b; q= u, s)$ systems, the quarks, $Q$ and $s$, have no contributions to the total isospin, which is only determined by the flavor of $\bar{q}$. There are six types of the flavor wave functions in the diquark-antidiquark picture,
\begin{subequations}
\begin{align}
\chi^{d1}&=(QQ\bar{s}\bar{s})_{1324}, \\
\chi^{d2}&=(QQ\bar{u}\bar{s})_{1324}, \\
\chi^{d3}&=(QQ\bar{s}\bar{u})_{1324}, \\
\chi^{d4}&=(bc\bar{s}\bar{s})_{1324}, \\
\chi^{d5}&=(bc\bar{u}\bar{s})_{1324}, \\
\chi^{d6}&=(bc\bar{s}\bar{u})_{1324}.
\end{align}
\end{subequations}
For the meson-meson picture, the flavor wave functions can be expressed as,
\begin{subequations}
\begin{align}
\chi^{t1}&=(Q\bar{s}Q\bar{s})_{1234}, \\
\chi^{t2}&=(Q\bar{u}Q\bar{s})_{1234}, \\
\chi^{t3}&=(Q\bar{s}Q\bar{u})_{1234}, \\
\chi^{t4}&=(b\bar{s}c\bar{s})_{1234}, \\
\chi^{t5}&=(b\bar{u}c\bar{s})_{1234}, \\
\chi^{t6}&=(b\bar{s}c\bar{u})_{1234}.
\end{align}
\end{subequations}
It needs to note that the subscripts of the wave functions for color and flavor stand for the indices of particles. Taking all degrees of freedom into account, we give the spin-flavor-color basis of $S\mbox{-}$wave $QQ\bar{q}\bar{s}~(Q=c, b; q= u, s)$ systems constrained by Pauli principle in Table \ref{wavefunctions1}. Finally, the complete wave function $\Psi_{IJ}^{M_IM_J}$ is obtained by coupling the orbital and spin, flavor, color wave functions. Actually, a real physical state should be the mixture of these basis with the same quantum numbers $I(J^P)$. And we can obtain the ground-state masses and the eigenvectors of the double-heavy strange tetraquarks by solving the Schr\"{o}dinger equation with the Rayleigh-Ritz variational principle,
\begin{equation}
    H \, \Psi^{\,M_IM_J}_{IJ}=E^{IJ} \Psi^{\,M_IM_J}_{IJ}.
\end{equation}

\begin{table*}[!t]
\linespread{1.2}
\begin{center}
\renewcommand\tabcolsep{1.0pt} % 调整表格列间的宽度
\caption{ \label{wavefunctions1} The spin-flavor-color basis of $S$-wave $QQ\bar{q}\bar{s}~(Q=c, b; q= u, s)$ systems constraint of Pauli principle. For simplicity, in the table, we use the superscripts of the $\chi$ to represent the wave functions, such as, $\chi_{00=0\oplus0}^{\sigma1}=\sigma1$, $\chi^{c1}_{1 \otimes 1} = c1$, $\chi^{d1}=d1$.}
\begin{tabular}{ccccccccccccc}
\hline\hline\noalign{\smallskip}
&\multicolumn{3}{c}{$QQ\bar{s}\bar{s}$} &\multicolumn{3}{c}{$QQ\bar{u}\bar{s}$} &\multicolumn{3}{c}{$bc\bar{s}\bar{s}$} &\multicolumn{3}{c}{$bc\bar{u}\bar{s}$} \\
\hline
&$0(0^+)$ &$0(1^+)$  &$0(2^+)$      &$\frac{1}{2}(0^+)$       &$\frac{1}{2}(1^+)$        &$\frac{1}{2}(2^+)$   &$0(0^+)$   &$0(1^+)$ &$0(2^+)$  &$\frac{1}{2}(0^+)$ &$\frac{1}{2}(1^+)$ &$\frac{1}{2}(2^+)$\\
&$(\sigma1,t1,c1)$&$(\sigma3,t1,c1)$&$(\sigma6,t1,c1)$&$(\sigma1,t2,c1)$&$(\sigma3,t2,c1)$&$(\sigma6,t2,c1)$&$(\sigma1,t3,c1)$&$(\sigma3,t3,c1)$&$(\sigma6,t3,c1)$&$(\sigma1,t5,c1)$&$(\sigma3,t5,c1)$&$(\sigma6,t5,c1)$\\
&$(\sigma1,t1,c2)$&$(\sigma3,t1,c2)$&$(\sigma6,t1,c2)$&$(\sigma1,t2,c2)$&$(\sigma3,t2,c2)$&$(\sigma6,t2,c2)$&$(\sigma1,t3,c2)$&$(\sigma3,t3,c2)$&$(\sigma6,t3,c2)$&$(\sigma1,t5,c2)$&$(\sigma3,t5,c2)$&$(\sigma6,t5,c2)$\\
&$(\sigma2,t1,c1)$&$(\sigma4,t1,c1)$&$(\sigma6,d1,c3)$&$(\sigma2,t2,c1)$&$(\sigma4,t2,c1)$&$(\sigma6,t3,c1)$&$(\sigma2,t3,c1)$&$(\sigma4,t3,c1)$&$(\sigma6,d3,c3)$&$(\sigma2,t5,c1)$&$(\sigma4,t5,c1)$&$(\sigma6,t6,c1)$\\
&$(\sigma2,t1,c2)$&$(\sigma4,t1,c2)$&                 &$(\sigma2,t2,c2)$&$(\sigma4,t2,c2)$&$(\sigma6,t3,c2)$&$(\sigma2,t3,c2)$&$(\sigma4,t3,c2)$&                 &$(\sigma2,t5,c2)$&$(\sigma4,t5,c2)$&$(\sigma6,t6,c2)$\\
&$(\sigma1,d1,c4)$&$(\sigma5,d1,c3)$&                 &$(\sigma1,t3,c1)$&$(\sigma5,t2,c1)$&$(\sigma6,d2,c3)$&$(\sigma1,d3,c4)$&$(\sigma5,t3,c1)$&                 &$(\sigma1,t6,c1)$&$(\sigma5,t5,c1)$&$(\sigma6,d5,c3)$\\
&$(\sigma2,d1,c3)$&                 &                 &$(\sigma1,t3,c2)$&$(\sigma5,t2,c2)$&$(\sigma6,d3,c3)$&$(\sigma2,d3,c3)$&$(\sigma5,t3,c2)$&                 &$(\sigma1,t6,c2)$&$(\sigma5,t5,c2)$&$(\sigma6,d5,c4)$\\
&                  &                &                 &$(\sigma2,t3,c1)$&$(\sigma3,t3,c1)$&                 &                 &$(\sigma3,d3,c3)$&                 &$(\sigma2,t6,c1)$&$(\sigma3,t6,c1)$&$(\sigma6,d6,c3)$\\
&                  &                &                 &$(\sigma2,t3,c2)$&$(\sigma3,t3,c2)$&                 &                 &$(\sigma4,d3,c4)$&                 &$(\sigma2,t6,c2)$&$(\sigma3,t6,c2)$&$(\sigma6,d6,c4)$\\
&                  &                &                 &$(\sigma1,d2,c4)$&$(\sigma4,t3,c1)$&                 &                 &$(\sigma5,d3,c3)$&                 &$(\sigma1,d5,c3)$&$(\sigma4,t6,c1)$&                 \\
&                  &                &                 &$(\sigma2,d2,c3)$&$(\sigma4,t3,c2)$&                 &                 &                 &                 &$(\sigma1,d5,c4)$&$(\sigma4,t6,c2)$&                 \\
&                  &                &                 &$(\sigma1,d3,c4)$&$(\sigma5,t3,c1)$&                 &                 &                 &                 &$(\sigma2,d5,c3)$&$(\sigma5,t6,c1)$&                 \\
&                  &                &                 &$(\sigma2,d3,c3)$&$(\sigma5,t3,c2)$&                 &                 &                 &                 &$(\sigma2,d5,c4)$&$(\sigma5,t6,c2)$&                 \\
&                  &                &                 &                 &$(\sigma3,d2,c4)$&                 &                 &                 &                 &$(\sigma1,d6,c3)$&$(\sigma3,d5,c3)$&                 \\
&                  &                &                 &                 &$(\sigma4,d2,c3)$&                 &                 &                 &                 &$(\sigma1,d6,c4)$&$(\sigma3,d5,c4)$&                 \\
&                  &                &                 &                 &$(\sigma5,d2,c3)$&                 &                 &                 &                 &$(\sigma2,d6,c3)$&$(\sigma4,d5,c3)$&                 \\
&                  &                &                 &                 &$(\sigma3,d3,c4)$&                 &                 &                 &                 &$(\sigma2,d6,c4)$&$(\sigma4,d5,c4)$&                 \\
&                  &                &                 &                 &$(\sigma4,d3,c3)$&                 &                 &                 &                 &                 &$(\sigma5,d5,c3)$&                 \\
&                  &                &                 &                 &$(\sigma5,d3,c3)$&                 &                 &                 &                 &                 &$(\sigma5,d5,c4)$&                 \\
&                  &                &                 &                 &                 &                 &                 &                 &                 &                 &$(\sigma3,d6,c3)$&                  \\
&                  &                &                 &                 &                 &                 &                 &                 &                 &                 &$(\sigma3,d6,c4)$&                  \\
&                  &                &                 &                 &                 &                 &                 &                 &                 &                 &$(\sigma4,d6,c3)$&                  \\
&                  &                &                 &                 &                 &                 &                 &                 &                 &                 &$(\sigma4,d6,c4)$&                  \\
&                  &                &                 &                 &                 &                 &                 &                 &                 &                 &$(\sigma5,d6,c3)$&                  \\
&                  &                &                 &                 &                 &                 &                 &                 &                 &                 &$(\sigma5,d6,c4)$&                  \\
\hline\hline
\end{tabular}
\end{center}
\end{table*}
%%%%%%%%%%%%%%%%%%%%%%%%%%%%%%%%%%%%%%%%%%%%%%%%%%%%%%%%%%%%%%%%%%%%%%%%%%%%%%%%%%%%%%%%%%%%%%%%%%%%%%%%%%%%%%%%%%%%%%%

\section{Numerical Results}
\label{results}
In the framework of the chiral quark model, we present a systematic calculation about the double-heavy tetraquarks with strange flavor $QQ\bar{q}\bar{s}~(Q=c, b; q= u, s)$. Through solving the four-body Schr\"{o}dinger equation with GEM, the converged numerical ground-state masses are obtained. As we all know, a system always tends to take the position with the lowest energy. For a four-quark system, there is a stable configuration with two well separated mesons, the threshold we called, which plays an important role here. A tetraquark state should be stable against strong interaction
if its energy lies below all possible two-meson thresholds, and the decay must be weak or electromagnetic interaction. If the state has the energy higher than the sum of the masses of two mesons, the tetraquark state can fall apart into two mesons via strong interaction. Resonance states may exist. In the following, we will present our numerical results of looking for possible bound states and resonance states for $QQ\bar{q}\bar{s}~(Q=c, b; q= u, s)$.

\subsection{Looking for bound states}
In this subsection, we try to find the promising stable bound states for $QQ\bar{q}\bar{s}~(Q=c, b; q= u, s)$. The chiral quark model predictions on the lowest energies of the double-heavy strange tetraquark states with a set of given $I(J^P)$ are presented in Table \ref{boundstates}. In the table, $E_{\rm{MM}}$ and $E_{\rm{DA}}$ represents the ground-state mass just only in the pure meson-meson~(MM) picture and the diquark-antidiquark~(DA) picture, respectively. $E_{cc}$ is the ground-state energy by considering the coupling of the MM and the DA pictures. We express the theoretical lowest threshold of the double-heavy strange tetraquark as $E_{\rm{Theo}}$, with $E_{\rm{Theo}}=$Mass(meson1)+Mass(meson2). The relevant masses of mesons can be read in Table \ref{mesonspectra}. The binding energy of the states can be therefore defined as
\begin{equation}
E_{\rm{B}}=E_{cc}-E_{\rm{Theo}}.
\end{equation}

In the table, we find that the energies are always lower in the MM picture than those in the DA picture except $bb\bar{u}\bar{s}$ state with $\frac{1}{2}(1^+)$, and the coupling of the two pictures are small for the ground state for the double-heavy strange tetraquarks except $cc\bar{u}\bar{s}$ and $bb\bar{u}\bar{s}$ states with $\frac{1}{2}(1^+)$. In our previous work about $cc\bar{u}\bar{d}$ and $bb\bar{u}\bar{d}$ tetraquarks, we also obtained the lower energies for the DA picture and the coupling of the two pictures can not be neglected either\cite{Chen:2021tnn}. In the last column in Table \ref{boundstates}, we find two promising stable bound states. For the state $cc\bar{u}\bar{s}$ with $\frac{1}{2}(1^+)$, the chiral quark model indicates that the binding energy is about 7.0 MeV, which is well consistent with the lattice QCD result with 7.7 MeV binding energy \cite{Junnarkar:2018twb}. For $bb\bar{u}\bar{s}$ with $\frac{1}{2}(1^+)$, it has the binding energy 104.4 MeV. For the other double-heavy strange tetraquark states in Table \ref{boundstates}, they all lie above the corresponding lowest thresholds within the chiral quark model.

In Table \ref{contributions2}, the contributions of each potential in the system Hamiltonian of $cc\bar{u}\bar{s}$ and $bb\bar{u}\bar{s}$ system with $I(J^P)=\frac{1}{2}(1^+)$ are given for comparisons. From the table, we can easily found that for $bb\bar{u}\bar{s}$ system, the $K$ meson exchange is much larger than the $cc\bar{u}\bar{s}$ system and plays a role in obtaining the deeper bound energy. Because the larger mass of $b$ quark than $c$ quark, the relative kinetic energy will diminish. The repulsion of the system will decrease, and attraction will be predominant. Original stable state is broken, which leads to the smaller distance between quarks. For $K$ meson exchange, it will become larger with the decrease of the quark distance.

In our previous work about $T_{cc}$ states with $0(1^+)$, we obtained a deep bound state if all the structures, meson-meson and diquark-antidiquark, are considered. And the model can get a very shallow bound state, which is very consistent with the experimental data, if the structure is limited to meson-meson one. Now when replace $d$ quark with heavier $s$ quark, we get the different conclusions, we cannot find the bound states in the pure meson-meson structures or the diquark-antidiquark structures, but a shallow bound state with binding energy -7.0 MeV is obtained in the calculations of structures coupling. It would perhaps helpful to trace the appearance of deep bound state in nonstrange systems (equivalently the absence of deep bound state in the strange system). So in order to understand the difference of the binding energy of $T_{cc}$ and $cc\bar{u}\bar{s}$, we give the contributions of each potential in the system Hamiltonian by considering the couplings of the meson-meson and diquark-antidiquark structures in Table \ref{contributions3}. We can easily find that for $T_{cc}$, $\pi$ meson exchange plays an important role in getting a bound state. In comparison, for $cc\bar{u}\bar{s}$ system, the kaon exchange is weaker than the pion exchange because of $s$ quark's larger mass.

In the Table \ref{diquark}, we give the binding energies ($\Delta E$) of the diquark $\bar{u}\bar{d}$ and $\bar{u}\bar{s}$, as well as the contributions of each potential in the system Hamiltonian. The second and third column tells us that
$\bar{u}\bar{d}$ is much easier to be the bound state than $\bar{u}\bar{s}$. If we take the mass of $s$ quark to be that of $d$ quark, we get the deeper bound energy due to the stronger kaon exchange in the fourth column. If we take the mass of $K$ to be that of $\pi$, the kaon exchange doesn't seem to change a lot. By tuning the $m_s$ and $m_K$, we find that the main reason for the shallow bound state of the  $cc\bar{u}\bar{s}$ system coming from the heavier mass of $s$ quark.

What's more, in Table \ref{compare}, we give the binding energy of the state $bb\bar{u}\bar{s}$ with $\frac{1}{2}(1^+)$ within various theoretical methods for comparisons. In the table, most of the work get the bound state for this state, but the binding energy is different from each other. In General, the value is about 10 $\sim$ 50 MeV in the quark model~\cite{Deng:2018kly,Eichten:2017ffp,Park:2018wjk,Silvestre-Brac:1993zem,Meng:2020knc,Luo:2017eub,Noh:2021lqs}, 100 MeV in the lattice QCD calculations~\cite{Francis:2016hui,Junnarkar:2018twb,Pflaumer:2021ong,Braaten:2020nwp}, 180 $\sim$ 500 MeV in the QCD sum rules \cite{Agaev:2020zag, Tang:2019nwv}. In the references \cite{Ebert:2007rn,Lu:2020rog}, the authors get no bound states for $bb\bar{u}\bar{s}$ with $\frac{1}{2}(1^+)$ in the relativistic quark model. Our results for $bb\bar{u}\bar{s}$ with $\frac{1}{2}(1^+)$ state is also well consistent with the lattice QCD calculations ~\cite{Francis:2016hui,Junnarkar:2018twb,Pflaumer:2021ong,Braaten:2020nwp}.

\begin{table*}[!t]
\linespread{1.2}
\begin{center}
\renewcommand\tabcolsep{15.0pt} % 调整表格列间的宽度
\caption{ \label{boundstates} The ground-state mass of double-heavy strange tetraquarks $QQ\bar{q}\bar{s}~(Q=c, b; q= u, s)$, masses unit in MeV.}
\begin{tabular}{ccccccc}
\hline\hline\noalign{\smallskip}
Flavor             &$I(J^P)$ &$E_{\rm{MM}}$ &$E_{\rm{DA}}$ &$E_{cc}$  &$E_{\rm{Theo}}$ &$E_{\rm{B}}$ \\
\hline
$cc\bar{s}\bar{s}$ &$0(0^+)$ &3906.5      &4217.4        &3906.5   &3900.4($D_s^+D_s^+$)         &6.1 \\
                   &$0(1^+)$ &4033.8      &4281.9        &4033.8   &4030.1($D_s^+D_s^{*+}$)         &3.7 \\
                   &$0(2^+)$ &4161.2      &4345.5        &4161.0   &4159.8($D_s^{*+}D_s^{*+}$)         &1.2 \\
\hline
$bb\bar{s}\bar{s}$ &$0(0^+)$ &10736.5     &10891.0       &10736.5  &10735.4($\bar{B}_s^0\bar{B}_s^0$)         &1.1  \\
                   &$0(1^+)$ &10778.8     &10916.4       &10778.8  &10777.9($\bar{B}_s^0\bar{B}_s^{*0}$)       &0.9  \\
                   &$0(2^+)$ &10820.9     &10943.9       &10820.9  &10820.4($\bar{B}_s^{*0}\bar{B}_s^{*0}$)     &0.5  \\
\hline
$cc\bar{u}\bar{s}$ &$\frac{1}{2}(0^+)$    &3814.3  &4193.5    &3814.2  &3812.8($D^0D_s^+$) &1.4 \\
                   &$\frac{1}{2}(1^+)$    &3933.0  &4020.9    &3923.8  &3930.8($D^{*0}D_s^+$) &-7.0 \\
                   &$\frac{1}{2}(2^+)$    &4063.0  &4289.1    &4062.9  &4060.5($D^{*0}D_s^{*+}$) &2.4 \\
\hline
$bb\bar{u}\bar{s}$ &$\frac{1}{2}(0^+)$    &10649.2 &10834.9   &10649.2 &10648.5($B^-\bar{B}_s^0$) &0.7 \\
                   &$\frac{1}{2}(1^+)$    &10687.8 &10615.9   &10582.9 &10687.3($B^{*-}\bar{B}_s^0$) &-104.4   \\
                   &$\frac{1}{2}(2^+)$    &10730.9 &10877.5   &10730.9 &10729.8($B^{*-}\bar{B}_s^{*0}$) &1.1 \\
\hline
$bc\bar{s}\bar{s}$ &$0(0^+)$ &7322.1 &7650.7 &7322.1 &7317.9($\bar{B}_s^0D_s^+$) &4.2 \\
                   &$0(1^+)$ &7364.3 &7662.2 &7364.3 &7360.4($\bar{B}_s^{*0}D_s^+$) &3.9 \\
                   &$0(2^+)$ &7491.1 &7696.3  &7491.1 &7490.1($\bar{B}_s^{*0}D_s^{*+}$) &1.0 \\
\hline
$bc\bar{u}\bar{s}$ &$\frac{1}{2}(0^+)$ &7232.0 &7317.8 &7231.6 &7230.3($\bar{B}_s^0D^0$) &1.3 \\
                   &$\frac{1}{2}(1^+)$ &7274.1 &7334.2 &7272.4 &7269.8($B^{*-}D_s^{+}$) &2.6 \\
                   &$\frac{1}{2}(2^+)$ &7392.2 &7582.0 &7392.1 &7390.8($\bar{B}_s^{*0}D^{*0}$) &1.3 \\
\hline\hline
\end{tabular}
\end{center}
\end{table*}

\begin{table}[!t]
\linespread{1.2}
\begin{center}
\renewcommand\tabcolsep{26.0pt} % 调整表格列间的宽度
\caption{ \label{contributions2} Contributions of each potential in the system Hamiltonian of $cc\bar{u}\bar{s}$ and $bb\bar{u}\bar{s}$ system with $I(J^P)=\frac{1}{2}(1^+)$, when considering the coupling of the meson-meson structure and diquark-antidiquark structure. $K_1$, $K_2$, $K_3$ represents the inner kinetic energy and relative kinetic energy between two clusters. $\Delta E$ is the binding energy of the system. (in unit of MeV).}
\begin{tabular}{ccc}
\hline\hline\noalign{\smallskip}
                       &$cc\bar{u}\bar{s}$   &$bb\bar{u}\bar{s}$ \\
\hline
$K_1$                  &434.6     &546.0     \\
$K_2$                  &384.0     &394.2     \\
$K_3$                  &202.7     &166.8     \\
$V^{\rm{C}}$           &-457.5     &-507.5     \\
$V^{\rm{Coul}}$        &-648.8     &-658.1     \\
$V^{\rm{CMI}}$         &-204.3     &-212.9     \\
$V^{\eta}$             &-17.0     &-40.5     \\
$V^{\pi}$              &0.0     &0.0     \\
$V^{\sigma}$           &-24.9     &-47.2    \\
$V^{K}$                &-74.9     &-178.1     \\
$E$                    &3923.8     &10582.9     \\
$\Delta E$             &-7.0     &-104.4     \\
\hline\hline
\end{tabular}
\end{center}
\end{table}

\begin{table}[!t]
\linespread{1.2}
\begin{center}
\renewcommand\tabcolsep{26.0pt} % 调整表格列间的宽度
\caption{ \label{contributions3} Contributions of each potential in the system Hamiltonian of $T_{cc}$ states with $I(J^P)=0(1^+)$ and $cc\bar{u}\bar{s}$ system with $I(J^P)=\frac{1}{2}(1^+)$, when considering the coupling of the meson-meson structures and diquark-antidiquark structures. $\Delta E$ is the binding energy of the state. (in unit of MeV). }
\begin{tabular}{ccc}
\hline\hline\noalign{\smallskip}
                   &$T_{cc}$    &$cc\bar{u}\bar{s}$    \\
\hline
$K_1$              &321.8       &434.6   \\
$K_2$              &929.7       &384.0           \\
$K_3$              &189.9       &202.7\\
$V^{\rm{C}}$       &-432.8      &-457.5        \\
$V^{\rm{Coul}}$    &-645.8      &-648.8   \\
$V^{\rm{CMI}}$      &-346.4     &-204.3         \\
$V^{\eta}$         &75.4        &-17.0\\
$V^{\pi}$          &-464.6      &0.0        \\
$V^{\sigma}$       &-48.8       &-24.9 \\
$V^{K}$            &0           &-74.9 \\
$E$                &3660.7      &3923.8\\
$\Delta E$         &-182.5      &-7.0          \\
\hline\hline
\end{tabular}
\end{center}
\end{table}

\begin{table*}[!t]
\linespread{1.2}
\begin{center}
\caption{ \label{diquark} The mass of the diquark $\bar{u}\bar{d}$ and $\bar{u}\bar{s}$. $\Delta E$ is the binding energy of the state. (in unit of MeV). }
\begin{tabular}{ccccc}
\hline\hline\noalign{\smallskip}
                   &$\bar{u}\bar{d}$    &\multicolumn{3}{c}{$\bar{u}\bar{s}$}   \\
                   &~~~~~~                    &~~~~~~$M_s=536,M_K=494$       &~~~~~~$M_s=313,M_K=494$ &~~~~~$M_s=536,M_K=139$ \\
\hline
%$K_1$              &0.7      &0.7       &0.7 &0.7                \\
$K_2$              &970.4    &686.9     &1133.8 &624.2        \\
%$K_3$              &1.1      &0.9       &1.1 &0.9         \\
$V^{\rm{C}}$       &-85.8    &-87.3     &-103 &-82.9          \\
$V^{\rm{Coul}}$    &-208.9   &-184.3    &-224 &-176.2              \\
$V^{\rm{CMI}}$     &-330.4   &-204.7    &-393 &-180.9               \\
$V^{\eta}$         &80.4     &-40.2     &-93.9 &-35.2               \\
$V^{\pi}$          &-495.3   &0         &0 &0             \\
$V^{\sigma}$       &-51.1    &-46.9     &-54.4 &-44.7               \\
$V^{K}$            &0        &-176.7    &-411.8 &-156.8          \\
%$E$                &507      &797.4     &479.8 &797.9          \\
$\Delta E$         &-120.7     &-53.2     &-146.3 &-52.5            \\
\hline\hline
\end{tabular}
\end{center}
\end{table*}

\begin{table}[!t]
%\linespread{1.2}
\begin{center}
\caption{ \label{compare} The stable tetraquark state $bb\bar{u}\bar{s}$ with quantum numbers $I(J^P)=\frac{1}{2}(1^+)$ in various methods, units in MeV.}
\begin{tabular}{cccccccccc}
\hline\hline\noalign{\smallskip}
States  &$I(J^P)$ &Ours &\cite{Deng:2018kly} &\cite{Eichten:2017ffp} &\cite{Francis:2016hui} &\cite{Park:2018wjk} &\cite{Silvestre-Brac:1993zem} &\cite{Meng:2020knc} &\cite{Junnarkar:2018twb} \\
$bb\bar{u}\bar{s}$    &$\frac{1}{2}(1^+)$   &-7.0     &-49 &-48 &-98 &-7 &-40 &-59 &-87  \\

&  &\cite{Pflaumer:2021ong} &\cite{Agaev:2020zag} &\cite{Braaten:2020nwp} &\cite{Luo:2017eub}  &\cite{Tang:2019nwv} &\cite{Noh:2021lqs} &\cite{Ebert:2007rn} &\cite{Lu:2020rog}                                        \\
&&-80 &-477 &-73 &-48 &-182  &-42 &14 &40           \\
\hline\hline
\end{tabular}
\end{center}
\end{table}

\subsection{Looking for resonance states}
\begin{figure}
\center{\includegraphics[width=7.0cm]{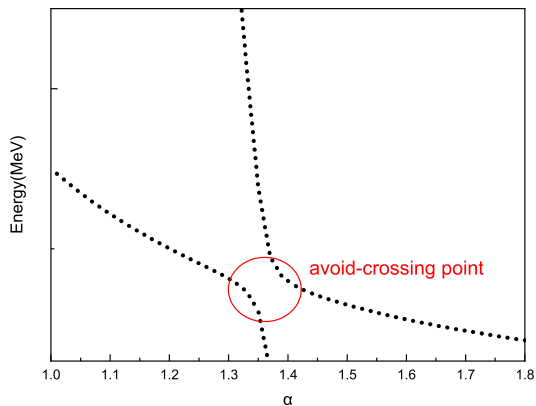}}
\caption{\label{avoid-crossing} Stabilization graph for the resonance.}
\end{figure}

\begin{figure*}
\center{\includegraphics[width=19.0cm]{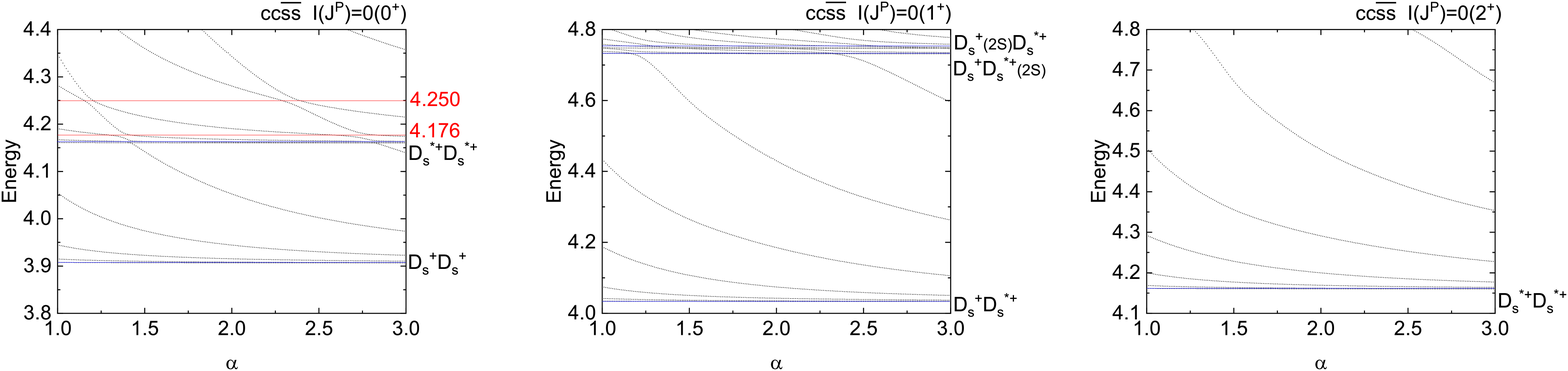}}
\caption{\label{ccss} The stabilization plots of the energies of $cc\bar{s}\bar{s}$ states with respect to the scaling factor $\alpha$, masses unit in GeV.}
\end{figure*}

\begin{figure*}
\center{\includegraphics[width=19.0cm]{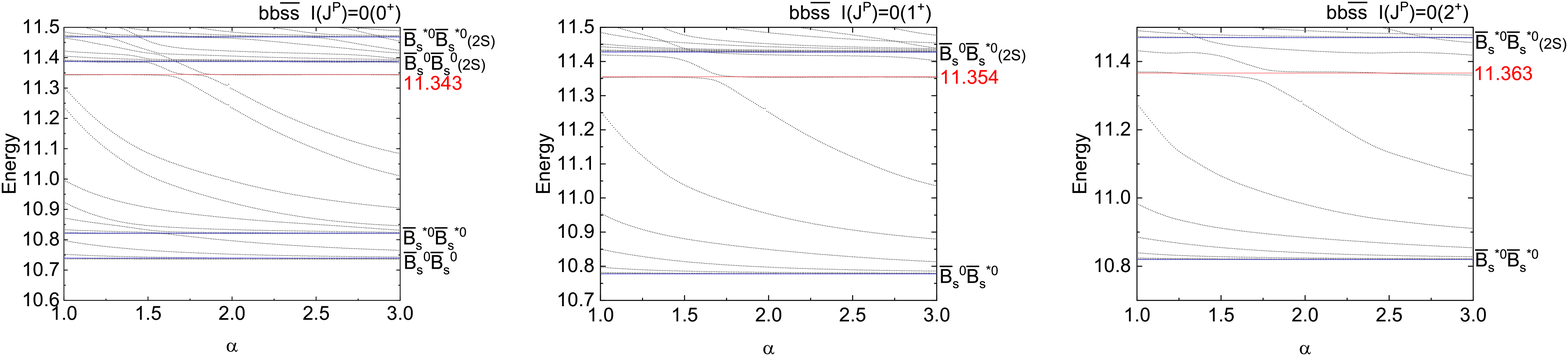}}
\caption{\label{bbss} The stabilization plots of the energies of $bb\bar{s}\bar{s}$ states with respect to the scaling factor $\alpha$, masses unit in GeV.}
\end{figure*}

\begin{figure*}
\center{\includegraphics[width=19.0cm]{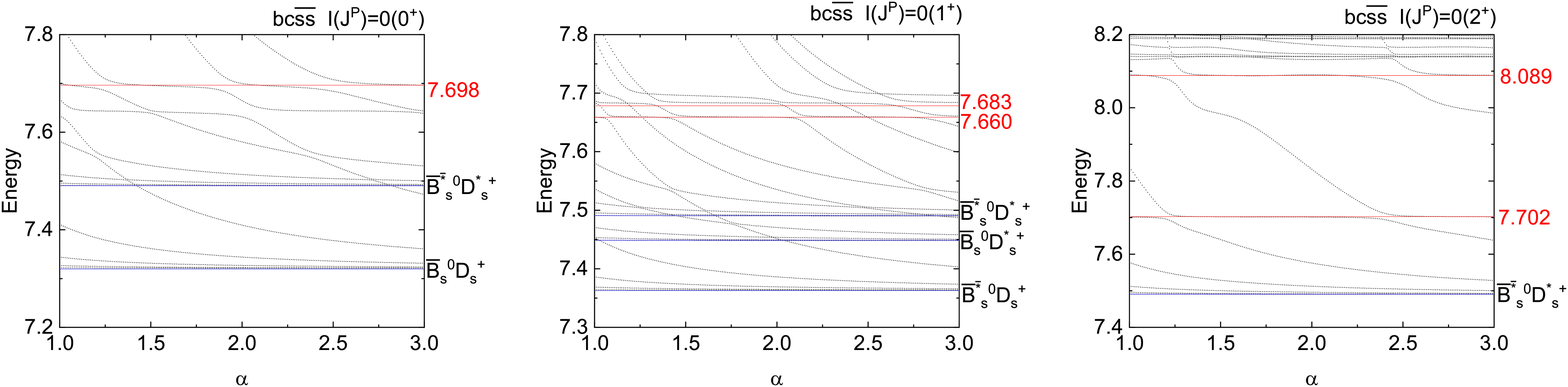}}
\caption{\label{bcss} The stabilization plots of the energies of $bc\bar{s}\bar{s}$ states with respect to the scaling factor $\alpha$, masses unit in GeV.}
\end{figure*}

\begin{figure*}
\center{\includegraphics[width=19.0cm]{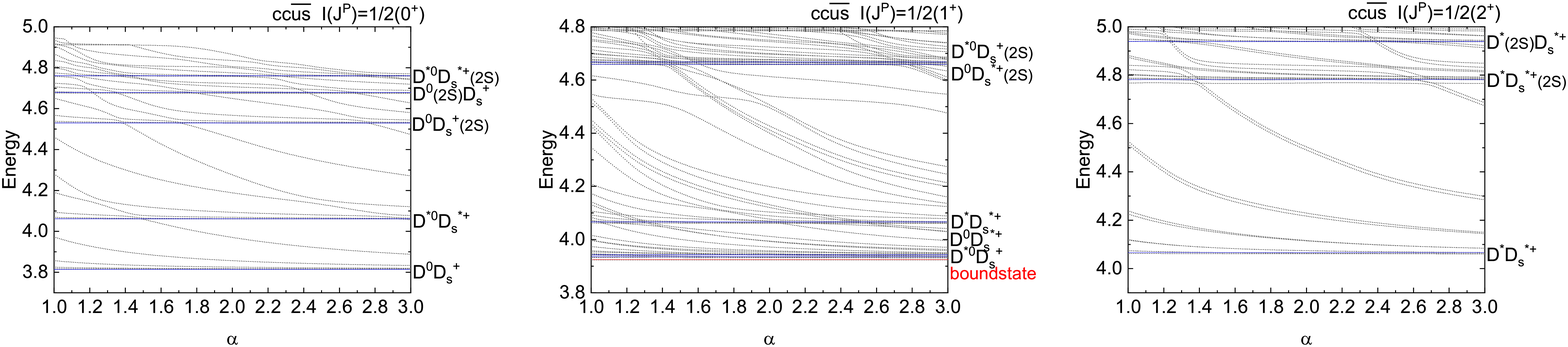}}
\caption{\label{ccus} The stabilization plots of the energies of $cc\bar{u}\bar{s}$ states with respect to the scaling factor $\alpha$, masses unit in GeV.}
\end{figure*}

\begin{figure*}
\center{\includegraphics[width=19.0cm]{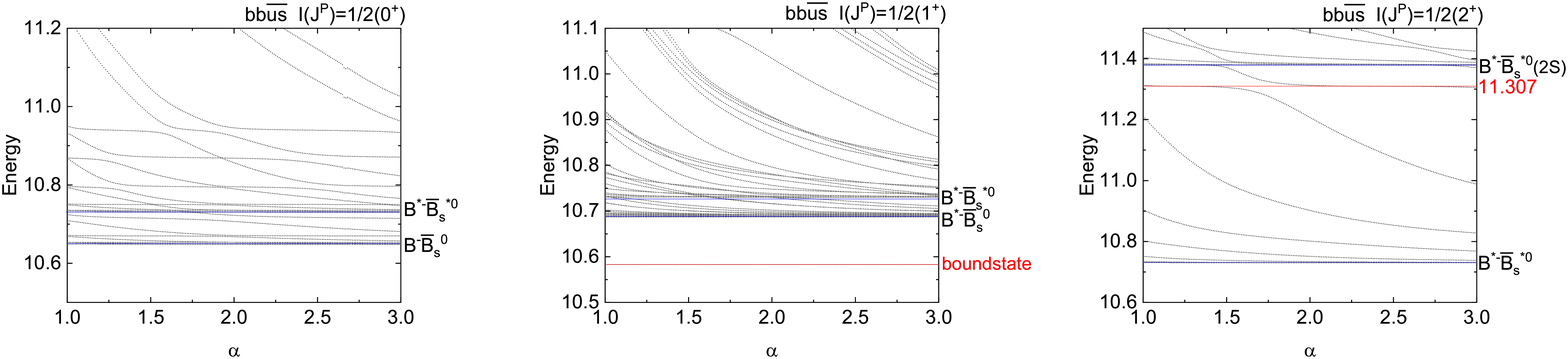}}
\caption{\label{bbus} The stabilization plots of the energies of $bb\bar{u}\bar{s}$ states with respect to the scaling factor $\alpha$, masses unit in GeV.}
\end{figure*}

\begin{figure*}
\center{\includegraphics[width=19.0cm]{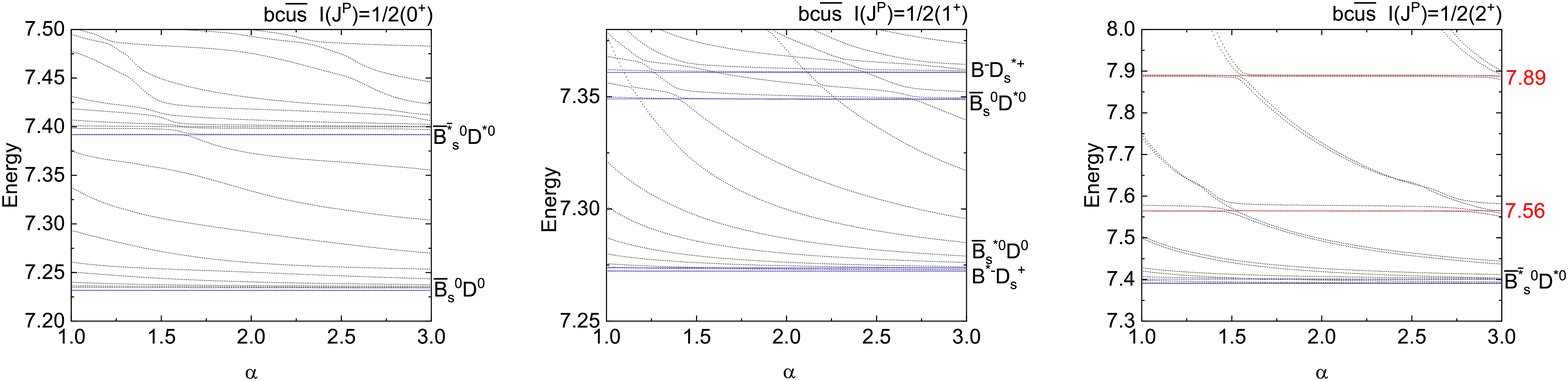}}
\caption{\label{bcus} The stabilization plots of the energies of $bc\bar{u}\bar{s}$ states with respect to the scaling factor $\alpha$, masses unit in GeV.}
\end{figure*}

Next we try to find possible resonance states for double-heavy strange tetraquarks $QQ\bar{q}\bar{s}~(Q=c, b; q= u, s)$. Herein, the real scaling method (RSM) \cite{rs1} is applied to find
the genuine resonances. In the RSM, the Gaussian range parameters $r_n$ in Eq. (\ref{gaussiansize}) just only for the meson-meson picture with the color singlet-singlet configuration are scaled as $r_n \rightarrow \alpha r_n$. In our calculations, $\alpha$ takes the values from 1.0 to 3.0. When $\alpha$ is varied, the distance of the two color-singlet mesons is scaled with $\alpha$. The energy eigenvalues of the scattering states will decrease as $\alpha$ increases. For the bound states, they will be shown as the straight lines, and resonance states will appear as an avoid-crossing structure like Fig. \ref{avoid-crossing}. Thus the $\alpha$ dependence of energy eigenvalues can be used to distinguish resonance states from scattering states. The results are demonstrated in Figs. \ref{ccss} to \ref{bcus}.

In Fig. \ref{ccss}, for the state $cc\bar{s}\bar{s}$ with $0(0^+)$, the blue lines represent the theoretical thresholds $D_s^+D_s^+$ and $D_s^{*+}D_s^{*+}$. The red lines are the two possible resonance states with mass 4176 and 4250 MeV. By analysing the components of the resonance states, we found that for the resonance with the energy 4176 MeV, $D_s^*D_s^*$ channel with the hidden color component takes up a larger proportion, to be $54\%$. For the resonance with the energy 4250 MeV, $D_sD_s$ channel with the hidden color component occupies larger proportion, to be $44\%$. Need to be noted that the wave function of resonance is not square-integrable and the proportions of different channels are defined in the sense of finite volume. The proportions of the resonances could be changed by varying the ``finite volume". In our work, we normalize the scattering state in the finite space occupied by the bound state (in the calculation without open channels involved), then the proportions of various components in the resonances will not depend on ``finite volume" anymore. The details can be seen in the reference \cite{finitevolume}. Besides, these proportions can reflect the strengths of the coupling between the resonances and the corresponding channels. For $0(1^+)$ and $0(2^+)$ states, we find no resonance states. In the reference \cite{Wu:2021rrc}, the authors also get a resonance state with the energy 4256 MeV for $0(0^+)$ in the quark delocalization color screening model (QDCSM), which is well consistent with our results.

In Fig. \ref{bbss}, for $bb\bar{s}\bar{s}$ states, we respectively find one resonance for $0(0^+)$, $0(1^+)$ and $0(2^+)$, with the mass 11343 MeV, 11354 MeV and 11363 MeV. The resonance mechanism of these resonances also comes from the hidden color structure since the hidden color component occupies a large proportion.

In Fig. \ref{bcss}, there are more fund of resonances for $bc\bar{s}\bar{s}$. For example, we find one resonance with energy 7698 MeV for $0(0^+)$, coming from the hidden color channel resonance mechanism. For $0(1^+)$, two possible resonances between 7.5 GeV to 7.7 GeV are found. One with the energy 7660 MeV comes from the diquark-antidiquark with $6\otimes\bar{6}$ color resonance mechanism, and the other with the energy 7683 MeV comes from the $B_s^*D_s$ hidden color channel resonance mechanism. For $0(2^+)$, there are two resonances between 7.7 GeV to 8.1 GeV, with the energy 7702 and 8089 MeV. They both come from the colorful channel resonance mechanism, with $B_s^*D_s^*$ hidden color $(71\%)$ and diquark-antidiquark with $3\otimes\bar{3}$ $(74\%)$, respectively.

For $cc\bar{u}\bar{s}$, in Fig. \ref{ccus}, there's no resonance state for $\frac{1}{2}(0^+)$, $\frac{1}{2}(1^+)$ and $\frac{1}{2}(2^+)$.

For $bb\bar{u}\bar{s}$ in Fig. \ref{bbus}, we also found no resonances for both $\frac{1}{2}(0^+)$ and $\frac{1}{2}(1^+)$, but there is one resonance with energy 11307 MeV for $\frac{1}{2}(2^+)$. With the lower energy than $B^*B_s^*(2S)$ channel, this resonance state may come from the bound state of the $B^*B_s^*(2S)$.

In Fig. \ref{bcus}, for $bc\bar{u}\bar{s}$ system, there are two resonance states for $\frac{1}{2}(2^+)$ with energy 7560 and 7890 MeV. They also come from the color resonance mechanism since the hidden color channel and the diquark-antidiquark channels occupy a large component.

Succinctly, we collect all the resonance states in Table \ref{resonances}.

\begin{table}[!t]
\linespread{1.2}
\begin{center}
\renewcommand\tabcolsep{15.0pt} % 调整表格列间的宽度
\caption{ \label{resonances} The possible resonance states of double-heavy strange tetraquarks $QQ\bar{q}\bar{s}~(Q=c, b; q= u, s)$, masses unit in GeV.}
\begin{tabular}{ccc}
\hline\hline\noalign{\smallskip}
Flavor             &$I(J^P)$ &Resonance states \\
\hline
$cc\bar{s}\bar{s}$ &$0(0^+)$  & 4.176 \quad 4.250\\
                   &$0(1^+)$  & ...\\
                   &$0(2^+)$  & ...\\
\hline
$bb\bar{s}\bar{s}$ &$0(0^+)$  & 11.343\\
                   &$0(1^+)$  & 11.354\\
                   &$0(2^+)$  & 11.363\\
\hline
$cc\bar{u}\bar{s}$ &$\frac{1}{2}(0^+)$  & ... \\
                   &$\frac{1}{2}(1^+)$  & ...  \\
                   &$\frac{1}{2}(2^+)$  & ...  \\
\hline
$bb\bar{u}\bar{s}$ &$\frac{1}{2}(0^+)$  & ... \\
                   &$\frac{1}{2}(1^+)$  & ...  \\
                   &$\frac{1}{2}(2^+)$  & 11.307  \\
\hline
$bc\bar{s}\bar{s}$ &$0(0^+)$ & 7.698 \\
                   &$0(1^+)$ & 7.660 \quad 7.683\\
                   &$0(2^+)$ & 7.702 \quad 8.089 \\
\hline
$bc\bar{u}\bar{s}$ &$\frac{1}{2}(0^+)$ & ... \\
                   &$\frac{1}{2}(1^+)$ & ... \\
                   &$\frac{1}{2}(2^+)$ & 7.56 \quad 7.89 \\
\hline\hline
\end{tabular}
\end{center}
\end{table}

Further more, the decay widths of the resonances can be roughly estimated from the two-state crossing formula given in Reference \cite{rs1}, as
\begin{align}
\Gamma=4|V(\alpha)|\frac{\sqrt{|S_r||S_c|}}{|S_c-S_r|},
\end{align}
where, $V(\alpha)$ is the energy difference between the upper and lower branches at the avoid-crossing point with the same value $\alpha$. $S_r$ and $S_c$ are the slopes of the two crossing levels, continuum and resonance, respectively. It should be noted that the decay width is the partial strong two-body decay width to $S\mbox{-}$wave channels included in the calculation. For example, for $cc\bar{s}\bar{s}$ with $0(0^+)$, there are two resonance states with energy 4176 and 4250 MeV, which has the decay width of 18 and 24 MeV, respectively. This decay value is just the partial widths to $D_s^+D_s^+ $ and $D_s^{*+}D_s^{*+}$ channels. For the other resonance states in Figs. \ref{bbss} to \ref{bcus}, we found that the slopes of the resonance states lines approximately equal zero, so the partial two-body strong decay width may approximately be very small, and the decay width may come from the width of excited mesons themselves.

\section{Summary}
\label{summary}
We systematically study the doubly heavy tetraquark states $QQ\bar{q}\bar{s}$ with all possible quantum numbers constraint of the Pauli principle, where $Q=c,b; q=u, s$ within the framework of the chiral quark model.

In order to search for the possible stable states against strong interaction, the meson-meson picture, the diquark-antidiquark picture and the coupling calculations are considered respectively. The results predict that $cc\bar{u}\bar{s}$ and $bb\bar{u}\bar{s}$ tetraquarks with $\frac{1}{2}(1^+)$ are bound states against strong interactions, with the binding energy 7.0 and 104.4 MeV, respectively, which may be explored in the experiments in the near future. Meanwhile, some resonance states are also found with the real scaling method.

Until now, none of stable or resonant doubly heavy tetraquarks with strange flavor has been observed in experiments and therefore we need more investigations on their properties. The future experimental values will provide an opportunity to check the availability of the different theoretical models.

\section{Acknowledgment}
This work is partly supported by Natural Science Foundation of Jiangsu Province under Grant No. BK20221166, and the National Natural Science Foundation of China under Grants No. 11847145 and No. 11865019.

\end{document}